\documentclass[a4paper,11pt]{article}
\pdfoutput=1
\usepackage{jcappub}
\usepackage{amsfonts,amsmath,amssymb, changepage}
\usepackage{graphicx}
\usepackage{color}
\usepackage{natbib}
\usepackage{enumitem}
\newcommand{\rthis}[1]{\textcolor{black}{#1}}

\title{Bayesian Analysis of  time dependence of  DAMA annual modulation amplitude}

\author[a]{Srinikitha  Bhagvati}

\author[a,1]{and Shantanu Desai \note{Corresponding author.} }

\affiliation[a]{ Department of Physics, Indian Institute of Technology, Hyderabad, Telangana-502285, India}

\emailAdd{ep18btech11003@iith.ac.in}
\emailAdd{shntn05@gmail.com}

\abstract{We implement  a test of the variability of the per-cycle annual modulation amplitude    in the different phases of the DAMA/LIBRA experiment using Bayesian model comparison. Using frequentist methods,  a previous study~\cite{Kelso_2018}  had demonstrated that the DAMA amplitudes  spanning over the DAMA/NaI and the first phase of the DAMA/LIBRA phases,  show a mild preference for  time-dependence in  multiple energy bins. With that motivation, we first show using Bayesian techniques that the  aforementioned data analyzed in ~\cite{Kelso_2018} show a moderate preference for exponentially varying amplitudes in the 2-5 and 2-6 keV energy intervals. We then carry out a  similar analysis  on the latest  modulation amplitudes released by the DAMA collaboration  from the first two phases of the upgraded DAMA/LIBRA  experiment. We also analyze the single-hit residual rates released by the DAMA collaboration to further look for any possible time-dependency. However, we  do not find any evidence for variability of either of the two datasets by using Bayesian model selection. All our analysis codes and datasets have been made publicly available. }
\begin{document}
\maketitle
\flushbottom


\section{Introduction}
The identity of  cold dark matter  is one of the most profound unsolved mysteries in Cosmology and fundamental Physics. For close to 90 years, we have known  that most of the mass of our galaxy and the rest of the universe is composed of dark matter~\cite{Hooper}. Further observations have shown that it is non-baryonic, collisionless, and that it decouples while moving at non-relativistic velocities~\cite{Kamionkowski,Silk}. Hence, the name ``cold dark matter''. The latest CMB and LSS results indicate that the total density of cold dark matter is about 25\%~\cite{Planck20}.


A large number of cold dark matter candidates have since been proposed, spanning many orders of magnitude in mass and  interaction strengths~\cite{Leszek,Feng}. Among those, the Weakly Interacting Massive Particle (WIMP) is one of the most well motivated and widely studied cold dark matter candidate. This is because any elementary particle which is a thermal relic from the Big-Bang having electro-weak scale interactions with ordinary matter satisfies all the properties needed for a cold dark matter candidate, such as the correct  relic abundance and  non-relativistic velocities at the time of decoupling~\cite{Weinberg,Goldberg,Dasgupta}.  

The detection of WIMPs using coherent elastic scattering with nuclei was proposed more than thirty years ago~\cite{Drukier84,Witten}. Since then, the experimental detection of WIMPs and measurements of their couplings with ordinary matter  have been the subject of a large number of experiments in various underground laboratories~\cite{Bauer}. The DAMA experiment is the only one among all these experiments so far to have claimed to detect dark matter.

The main basis for this claim by the DAMA collaboration is the  observation of an annual cosine-based modulation  in the residual single-hit count rates in the 2-6 keV energy intervals and thereabouts. This observed modulation has the correct phase (peak in June and minima in December) and amplitude that  is in accord with the theoretical predictions of dark matter halo models~\cite{Drukier,Freese88}. 
Since its inception in 1995, the statistical significance of the DAMA signal has been growing steadily with accumulated data over the different phases of the  experiment~\citep{DAMA03,Bernabei:2008yi,DAMA10,DAMA13,Bernabei_2018}.  The first data release showed only a $3\sigma$ significance for the annual modulation~\cite{DAMA00}. This significance increased to $6.3\sigma$ by the end of the first phase of the DAMA experiment (DAMA/NaI)~\cite{DAMA03}. The latest data release in 2018 shows a significance greater than $8\sigma$  across all the  energy intervals, which could contain a WIMP signal. The net significance in the 2-6 keV energy bin is equal to  $12.9\sigma$,  after combining all the accumulated data until  the end of phase 2 of the upgraded DAMA  experiment (DAMA/LIBRA)~\citep{Bernabei_2018}.

However, other direct detection experiments having different target materials  have failed to confirm this result and their upper limits completely rule out the allowed regions for WIMP mass and cross-section deduced from the DAMA annual modulation results \rthis{under the canonical assumptions for the dark matter}~\cite{Lux,PandaX,Aprile,CDEX}. Experiments having the same target material as DAMA (NaI) such as COSINE-100 and ANAIS-112, which  have  recently commenced operations  have also obtained null results~\cite{Cosine100,Anais112}. A large part of the DAMA parameter space has also been ruled out by some  indirect dark matter searches~\cite{Desai04}. Despite all this, it is important to scrutinize the DAMA dataset to see if it contains additional signatures which could help distinguish between the dark matter and an instrumental origin for the observed annual modulation.

In this work, we search for a time dependence of the DAMA best-fit annual modulation amplitude using Bayesian model comparison. This is a follow-up of a previous similar study done in  ~\cite{Kelso_2018}, who showed using frequentist model comparison techniques that the annual modulation amplitudes starting from the first ever phase of the DAMA experiment (starting in 1995)  upto 2013 showed hints of variability. We revisit this question using the latest DAMA data released in 2018 and use Bayesian model comparison techniques for this purpose.

The outline of this manuscript is as follows. Section \ref{sec:2} provides an introduction to  Bayesian model comparison techniques that will be used in this work. In Section \ref{sec:kelsorecap}, we give a brief overview of the analysis and results obtained for the time-dependence of the modulation amplitude in  ~\cite{Kelso_2018} as well as our re-analysis of the same data. Section \ref{sec:4} contains the description of the analysis and results for the model comparison done on the single-hit residual rates as well as the mean per-cycle modulation amplitude data released by the DAMA collaboration. Our conclusions are discussed in  Section \ref{sec:conclusions}.

\section{Introduction to Model Comparison Techniques}
\label{sec:2}
We provide a brief summary of  Bayesian model comparison techniques used in this work. This is the most robust method for model selection, surpassing both frequentist or information theoretical techniques~\cite{Sanjib}. More details on Bayesian model comparison can be found in various recent  reviews~\cite{astroml,Weller,Sanjib,Trotta} (and references therein). We have previously applied Bayesian model comparison techniques to a number of problems in Astrophysics and Cosmology, including the analysis of data from dark matter experiments such as DAMA, COSINE-100, and ANAIS-II~\cite{Krishak1,Krishak2,Krishak3,Krishak4,Haveesh,Rajdeep}. For more details, the reader can refer to ~\cite{astroml,Weller,Sanjib,Trotta} (and references therein). Previous applications of Bayesian model comparison while analyzing the  DAMA annual modulation can be found in ~\cite{Messina,Grilli}.

The starting point of Bayesian Model comparison for a model $M$ with respect to data $D$ is given by the Bayes Theorem. \rthis{The model M is usually the hypothesis which needs to be compared.} The Bayes theorem is given by:
\begin{equation}
        P(M|D) = \frac{P(D|M)P(M)}{P(D)},
        \label{eq:bayesthm}
    \end{equation} 
 where $P(M|D)$ is  the posterior probability and $P(D|M)$ the marginal likelihood, or sometimes  known as the Bayesian Evidence, \rthis{$P(M)$ is the prior probability on the model $M$ and $P(D)$ is the probability for the data $D$, which is usually never used in parameter estimation or model comparison}. This equation can be written as:
\begin{equation}
P(D|M)  \equiv \int P(D|M, \theta)P(\theta|M) \, d\theta
\label{eq:evid}
\end{equation}
where $\theta$ is the parameter vector  for the model $M$ and  $P(\theta|M)$ denotes the prior on $\theta$ for that model. 
In order to single out the preferred model between two models $M_1$ and $M_2$, we calculate the  odds ratio,  defined as the  ratio of their posterior probabilities. This odds ratio for the model $M_2$ over the model  $M_1$ is as follows: 
\begin{equation}
O_{21} = \frac{P(M_2|D)}{P(M_1|D)}
\end{equation}
Using Eq.~\ref{eq:bayesthm} and ~\ref{eq:evid}, this can be  simplified as follows:
\begin{equation} 
    O_{21} = \frac{E(M_2)P(M_2)}{E(M_1)P(M_1)} = B_{21}\frac{P(M_2)}{P(M_1)}\end{equation}
where the term $B_{21}$ is the Bayes Factor, given by the ratio of Bayesian Evidences for  the two models. For equal  prior probabilities for both the models, the Odds Ratio is identical to Bayes Factor. Therefore, we obtain:
\begin{equation} O_{21} = B_{21} = \frac{\int P(D|M_2, \theta_2)P(\theta_2|M_2) \, d\theta_2}{\int P(D|M_1, \theta_1)P(\theta_1|M_1) \, d\theta_1} \end{equation}
The Bayes factor is the quantity which is  used for Bayesian model comparison.
\rthis{We note that one key assumption in using the Bayes factor as a proxy for the odds ratio for model comparison, is that all the models compared have equal apriori probabilities.} 
The model with the larger value for Bayesian evidence will be considered as the preferred model. The  Jeffrey's scale is then used to qualitatively  assess its  significance~\cite{Trotta}.  A Bayes Factor $< 1$ indicates negative support for the model for $M_2$. A value greater than  10 indicates  moderate evidence for $M_2$, whereas a value  greater than  100 points to  decisive evidence.  
In this work, we calculate the Bayesian evidence for our models using {\tt Nestle}\footnote{\url{http://kylebarbary.com/nestle/}}, which is a python package  based on the nested sampling algorithm~\cite{Buchner}.

\section{Recap of Kelso18 and Re-analysis}
\label{sec:kelsorecap}
In their paper, Kelso et al~\cite{Kelso_2018} (K18, hereafter) studied the best-fit modulation amplitudes of the DAMA/NaI and DAMA/LIBRA Phase-I exposures. For this purpose, they analyzed the data released by the DAMA collaboration over the last 14 years starting in 1995 (covering both the  DAMA/NaI and DAMA/LIBRA phases) to explore the time dependence of the modulation amplitude using frequentist model comparison methods~\cite{Weller}. This would help discriminate between a dark matter versus an instrumental origin for the DAMA annual modulation signal.
They performed their analyses using  the mean-cycle amplitude data from the 2-4, 2-5, 2-6, 4-5, 4-6, and 5-6 keV energy ranges. 

The data used for the 2-4, 2-5, and 2-6 keV energy ranges was obtained from Figure 3 of Ref.~\cite{DAMA13}, provided by the DAMA collaboration. The mean time for each cycle is taken relative to January 1, 1995. The mean-cycle data for the higher energy ranges (4-5, 4-6, and 5-6 keV) were calculated from the efficiency-weighted average of the mean-cycle amplitudes in  the 2-4, 2-5 and 2-6 keV ranges. The main premise behind this is that the data from different energy ranges modulate with the same period and amplitude, but are statistically independent. 
Although this calculation is strictly correct only when the phase and period are fixed, K18 argued that this approximation is also valid when the phase and period vary, since the difference in the modulation amplitude between the fixed versus free period and phase case differs by less than 0.5$\sigma$. 

The best-fit modulation amplitudes for the DAMA/LIBRA Phase-I and DAMA/NaI exposures are calculated, along with the combined modulation amplitude for both the experiments. From the results listed in Table-I of K18, one can see by eye that the amplitudes decrease in every energy range, hinting towards  a decreasing trend in the amplitudes for the two  phases of  the DAMA experiment.  
By using multiple non-parametric  statistical tests, K18 concluded   that the null hypothesis is rejected at 2.4$\sigma$ in the 2-6 keV and $p-$values of 21\% and 5\% in the 2-4 and 2-5 keV energy bins. Therefore, they deduce that
one possible explanation for the results of these tests is that the modulation amplitude varies with time. 

To further quantify this, five parametric models for the time dependence of the  modulation amplitude were analyzed in K18: constant, broken constant with separate values for  DAMA/NaI data and DAMA/LIBRA Phase-I data, linearly time-dependent, exponentially time-dependent, and a broken exponential model with independent normalizations for NaI and LIBRA phases. 
Results from the goodness of fit test showed that all these models provided reasonable fits for data from all the energy ranges with the exception of the 5-6 keV range. 

The constant time-independent model was considered as the null hypothesis for model comparison. It was observed that all the energy ranges, except the 2-4 keV range, show a preference for time-dependence at the 2-3$\sigma$ significance level. The data in the 5-6 keV range show three departures at $>2.8\sigma$ from the mean value and do not fit any modulation hypothesis. Therefore, K18 point out that there could be  a problem for the data in 5-6 keV energy range. They conclude that all the time-varying models  in the 2-6 keV range are preferred over a constant amplitude model with a significance  between 2.3$\sigma$ to 2.8$\sigma$. However, a strong preference between the various time-dependent models could not be discerned using the data available until then.

\subsection{Re-analysis of K18 using Bayesian model selection}
As described above, K18 used frequentist methods to carry out model comparison between the time-independent constant model and various time-dependent models using the  per-cycle modulation amplitude data from DMA/NaI and the first phase of DAMA/LIBRA experiments. Before we move onto model comparison for the updated data from the second phase of the DAMA/LIBRA experiment, we redo the analysis using the aforementioned data analyzed in K18 using Bayesian model comparison for the linear and exponential models. The reason is that these models showed the largest significance when compared to a constant modulation amplitude and are representative of the simplest time-varying models. \rthis{However, we  also report the $\chi^2$/DOF for each of the models as a sanity check of the goodness of fit. A $\chi^2$ test is a statistical test used to compare the observed results of an experiment with the expected results.  The $\chi^2$ statistic is defined as $$\chi^2 = \sum \left(\frac{O_i - E_i}{\sigma_i}\right)^2$$ where $O_i$ and $E_i$ are the observed and expected values (for a particular model) respectively, and $\sigma_i$ indicates the error in the observed values. For all the results in this manuscript, the $\chi^2$ values are obtained from the comparison of the best-fit model to the observed data, while the $DOF$ corresponds to the number of degrees of freedom, calculated as the difference in number of data points and the number of free parameters.}

We use the data given in Tables 3 and 5 of K18 for the mean per-cycle modulation amplitudes and mean cycle times. The mean time for each cycle is taken relative to January 1, 1995. 
The  constant time-independent, linearly time-dependent and exponentially time-independent models for $S_m$ are the same as those in K18 and are given below:
\begin{align}
    &\text{Constant: } & S_{m} &= A_0 \label{eq:mean_const}\\
    &\text{Linear: } & S_{m} &= mt + c \label{eq:mean_linear}\\
    &\text{Exponential: } & S_{m} &= Ae^{-Bt} \label{eq:mean_exp}
\end{align}
 where $t$ denotes time, $A_0$, $m$, $c$, $A$ and $B$ are variable parameters used in the analysis.
Equations \ref{eq:mean_const}, \ref{eq:mean_linear} and \ref{eq:mean_exp} describe the three models used in our analysis. For evaluating the Bayes factors, we choose  uniform priors  for the parameters of these models, which  are listed in Column-I of Table \ref{table:bayes-priors1}.  The constant model is chosen as the null hypothesis. 

\begin{table}[]
\centering
\begin{tabular}{|c|c|c|}
\hline
 & Priors & Priors \\
Parameters & Per-cycle Modulation Amplitude analysis & Residual Rates Analysis\\ \hline
$A_0$  & [-0.05, 0.05]   &  [-0.05, 0.05] \\ 
$m$     & [-0.05, 0.05]   & [-0.05, 0.05] \\ 
$c$     & [-0.05, 0.05]   & [-0.05, 0.05] \\ 
$A$     & [-0.05, 0.05]   & [-0.05, 0.05] \\ 
$B$      & [-0.1 , 0.1]    & [-0.01, 0.01] \\ 
$\omega$ &  & [0.0015, 0.1281] \\ 
$t0$   &      & [0,$2\pi/\omega$] \\ \hline
\end{tabular}
\caption{The prior ranges chosen for the parameters used for all the analyses in this work.} 
\label{table:bayes-priors1}
\end{table}

Our results from Bayesian model selection are shown in Table~\ref{tab:kelso_bayes}. We find that none of the datasets show a decisive evidence (Bayes factor $>100$) for the linear or exponential model. The largest Bayes factors which we  obtained were for the exponential model in the 2-5 and 2-6 keV energy ranges of around 11 and 20, respectively. These point to ``moderate'' evidence for these models according to Jeffreys' scale~\cite{Trotta}. On the other hand, the Bayes factors for these models in the 4-5 keV and 4-6 keV energy range are still small, implying that there is no evidence for time variability in these energy ranges for these models.

 \begin{table}[]
     \centering
     \begin{tabular}{|c|c|c|c|}
        \cline{1-4}
        Data  &  Models & $\chi^2/DOF$ & Bayes Factors\\
        \cline{1-4}
        2-4 keV & Constant & 9.4/13 & - \\
         & Linear & 6.9/12 & 0.05\\
         & Exponential & 7.2/12 & 1.50\\
        \cline{1-4}
        2-5 keV & Constant & 13.6/13 & - \\
         & Linear & 6.4/12 & 0.36\\
         & Exponential & 7.5/12 & 11.71\\
        \cline{1-4}
        2-6 keV & Constant & 10.7/13 & -\\
         & Linear & 4.3/12 & 0.23\\
         & Exponential & 3.5/12 & 20.44\\
        \cline{1-4}
        4-6 keV & Constant & 22.4/13 & -\\
         & Linear & 19.0/12 & 0.05\\
         & Exponential & 15.2/12 & 5.49\\
        \cline{1-4}
        4-5 keV & Constant & 19.7/13 & -\\
         & Linear & 13.2/12 & 0.42\\
         & Exponential & 15.0/12 & 4.30\\
        \cline{1-4}
        5-6 keV & Constant & 40.0/13 & -\\
         & Linear & 39.8/12 & 0.02\\
         & Exponential & 35.1/12 & 1.28\\
        \cline{1-4} 
     \end{tabular}
     \caption{Results for the Bayesian \rthis{model comparison} analysis on the data from Tables 3 and 5 of K18. The mean time for each cycle is taken relative to January 1, 1995. The Bayes Factors are calculated for the linear and exponential models, while considering the constant model as the null hypothesis, and are listed in Column-IV of this table. We find that 2-5 keV and 2-6 keV energy intervals show moderate evidence for exponential variation of the amplitude. All other datasets are consistent with no variability. \rthis{We also list the $\chi^2$/dof for each of the models/energy ranges as a sanity check of the goodness of fit.} }
     \label{tab:kelso_bayes}
 \end{table}
\section{Analysis and Results}
\label{sec:4}

We now present our results on the search for possible time-dependence of the modulation amplitudes using the latest DAMA data released in 2018. We do this analysis using two independent datasets, as   further described in sub-sections \ref{sec:full} and \ref{sec:mean-value}, below.




\subsection{Analysis of single-hit residual rates from DAMA/LIBRA Phase-I and Phase-II} \label{sec:full}
\par
This sub-section entails the analysis of the data for the single-hit residual rates from the DAMA/LIBRA Phase-I and Phase-II experiments, obtained from \cite{Bernabei_2018, Bernabei:2008yi, DAMA13}. Data from the 2-4 keV and 2-5 keV energy ranges from the Phase-I of the DAMA/LIBRA experiment, 1-3 keV and 1-6 keV energy ranges in the Phase-II of the DAMA/LIBRA experiment, and 2-6 keV energy range from both Phase-I and Phase-II of the DAMA/LIBRA experiment are analyzed. We used the digitized DAMA data for these energy intervals from our previous work~\cite{Krishak1}. In the 2-6 keV energy range,  we also combined this data with that from DAMA/NaI experiment  (which was digitized from Ref \cite{Bernabei:2008yi}) and perform the same analysis. 
The total exposure for DAMA/NaI, the first phase of the DAMA experiment that ran for seven years, was $1.08 \times 10^5$ kg-days (0.29 ton-year). The exposure then increased over the the first phase of DAMA/LIBRA, the upgraded DAMA experiment, which registered a total exposure of $3.80 \times 10^5$ kg-days (1.04 ton-year) over the course of another seven years. The exposure further increased in the second phase of the DAMA/LIBRA experiment, registering a total exposure of $4.11 \times 10^5$ kg-days (1.13 ton-year). The total cumulative exposure, spanning DAMA/NaI and DAMA/LIBRA Phase-I and Phase-II, is 2.46 ton-year~\cite{Bernabei_2018}.

The models for the single-hit residual rate term, which we posit  for our  analysis are:
\begin{align}
&\text{Constant $S_m$:}&    f(t) &= A_0\cos{\omega (t - t0)} \label{eq:const}\\
&\text{Linear $S_m$:}& f(t) &= (mt + c)\cos{\omega (t - t0)} \label{eq:linear}\\
&\text{Exponential $S_m$:}& f(t) &= Ae^{-Bt}\cos{\omega (t - t0)} \label{eq:exponential}
\end{align}
where $t$ corresponds to time, $\omega$ and $t_0$ are the angular frequency and phase of the modulation, and $A_0$, $m$, $c$, $A$, $B$ represent the parameters used to model the time-variation of the amplitude $S_m$. Therefore, Eq.~\ref{eq:linear} and Eq.~\ref{eq:exponential} contain additional time-dependence on top of the well known cosine modulation.
We chose uniform priors on all these parameters, which are listed in Column-II of Table \ref{table:bayes-priors1}. The maximum value  for  the prior on $\omega$ is taken as $2\pi$ over the median difference in time data-points. Since the maximum period for which we can detect a potential modulation is half the total duration of the dataset, we used that to choose the minimum value for the prior on $\omega$.
 The ranges for the  priors on the other parameters have been chosen based on the maximum and minimum values for the residual count rates.

\begin{figure*}[t]
        \centering
        \includegraphics[width=\textwidth]{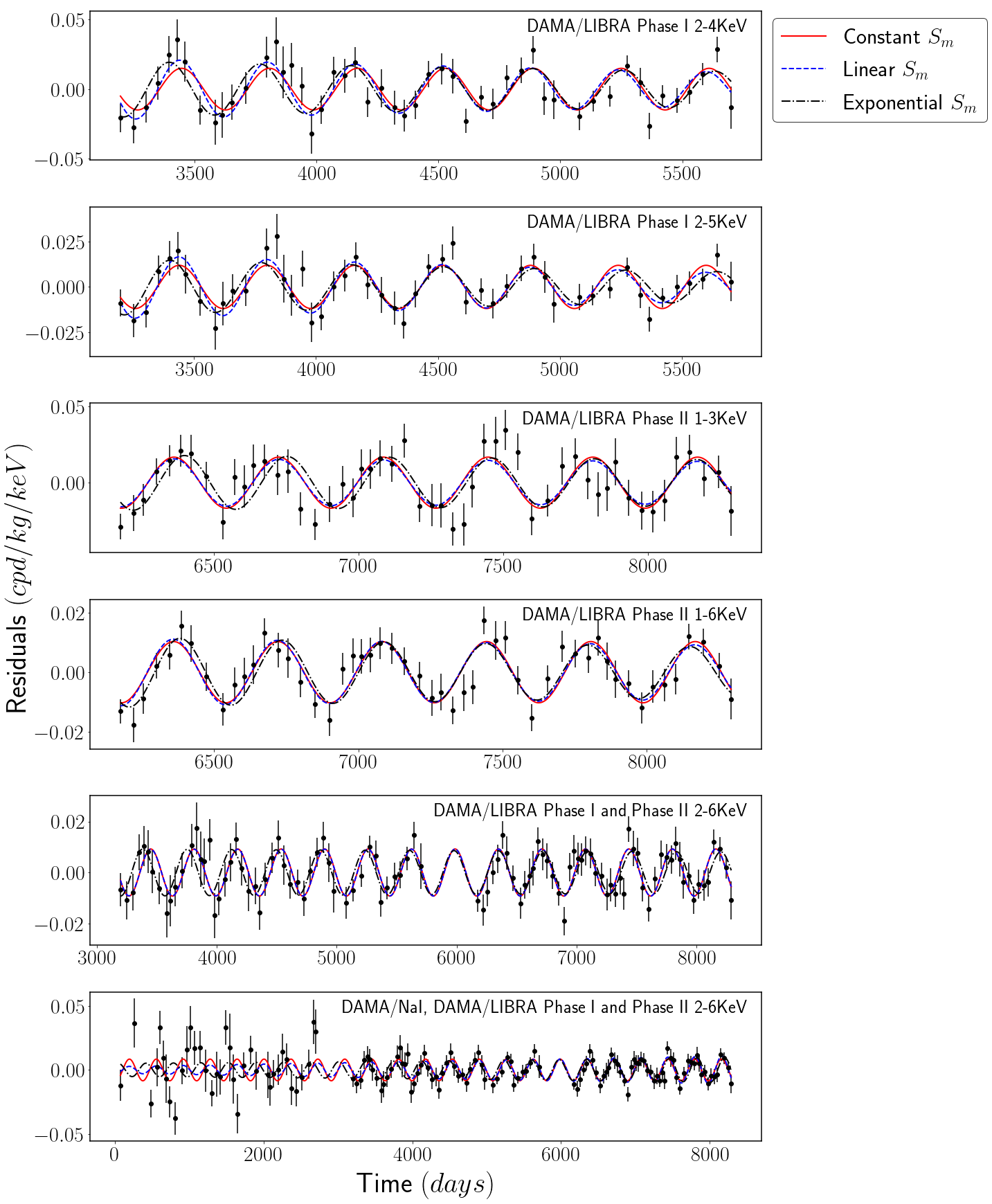}
        \caption{Plots of the best-fit curves of the constant, linearly dependent and exponentially dependent annual modulation amplitudes (on top of the expected cosine modulaton) from the analysis of residual rates described in Section \ref{sec:full}. The best-fits are calculated from the median of the posterior samples obtained from {\tt Nestle} used for Bayesian model comparison and the curves are plotted against the data released by the DAMA collaboration \cite{Bernabei_2018, Bernabei:2008yi, DAMA13}.}
        \label{fig:fulldata}
\end{figure*}

The best-fit plots for the constant, linearly dependent and exponentially dependent models for $S_m$ are shown in Figure \ref{fig:fulldata}.  The best-fit values used in the plot are calculated using the median of the samples and the curves are fitted against the data released by the DAMA collaboration \cite{Bernabei_2018, Bernabei:2008yi, DAMA13}.

\begin{table*}[]

\begin{tabular}{|c|c|c|c|}
\cline{1-4}
Data & Models & $\chi^2/DOF$ & Bayes Factors \\ 
\cline{1-4}
DAMA/LIBRA Phase-I: & Constant & $47.0/47$ & - \\
2-4 keV & Linear  & $44.1/46$ & $2.69\times 10^{-4}$\\
 & Exponential & $43.4/46$ & $4.77 \times 10^{-2}$\\ 
 \cline{1-4}
 
DAMA/LIBRA Phase-I: & Constant  & $38.3/47$ & -\\
2-5 keV & Linear  & $34.6/46$ & $2.87 \times 10^{-4}$\\
 & Exponential  & $34.6/46$ & $9.98 \times 10^{-3}$\\ 
 \cline{1-4}

DAMA/LIBRA Phase-II: & Constant & $52.3/49$ & - \\
1-3 keV & Linear & $51.7/48$ & $0.68 \times 10^{-4}$\\
 & Exponential  & $51.6/48$ &  $4.39 \times 10^{-3}$\\ 
 \cline{1-4}
 
DAMA/LIBRA Phase-II: & Constant  & $47.7/49$ & -\\
1-6 keV & Linear  & $49.2/48$ & $0.58 \times 10^{-4}$\\
 & Exponential & $47.0/48$ & $5.59 \times 10^{-3}$\\ 
 \cline{1-4}

DAMA/LIBRA Phases I and II: & Constant & $69.5/99$ & - \\
2-6 keV & Linear  & $69.5/98$ & $0.19 \times 10^{-4}$\\
 & Exponential & $69.4/98$ & $1.92 \times 10^{-6}$\\ 
 \cline{1-4}

DAMA/NaI, DAMA/LIBRA Phases I and II: & Constant  & $154.5/136$ & -  \\
2-6 keV & Linear  & $147.6/135$ & $1.49 \times 10^{-4}$\\
 & Exponential  & $149.8/135$ & $9.60 \times 10^{-5}$\\ 
 \cline{1-4}
 \end{tabular}
\caption{Summary of model comparison results for different models of the single-hit residual rates as discussed in Section \ref{sec:full}. \rthis{The $\chi^2$ values are calculated from the $\chi^2$ statistic by comparing the best-fit model to the data, and $DOF$ is taken as the difference in number of data points and the number of parameters.} The  Bayes factors for the linear (Equation \ref{eq:linear}) and exponentially (Equation \ref{eq:exponential}) varying modulation amplitudes have been calculated by taking the constant amplitude model (Equation \ref{eq:const}) as the null hypothesis.}
\label{table:results-1}
\end{table*}

The results for the Bayesian model comparison are summarized in  Table \ref{table:results-1} for the different energy ranges. In addition to the Bayes factors, we also list the $\chi^2$/DOF values. 
We find that both the constant amplitude models as well all  the time-varying amplitudes have $\chi^2$/dof close to 1. However,  the Bayes factor for all the time-varying amplitudes are less than or close to  one for all phases of the DAMA experiment. This shows that there is no evidence for linear or exponential time  variation of the modulation amplitude, when considering the residual count rates.


\subsection{Analysis of annual mean per-cycle modulation amplitude data} \label{sec:mean-value}

This section discusses the Bayesian analysis of mean per-cycle data for the modulation amplitude from Fig 14 of \cite{Bernabei_2018}. This is an update  of the analysis done in  K18  as discussed in Section \ref{sec:kelsorecap}. Although only the data from the  2-4, 2-5, and 2-6 keV energy ranges spanning the DAMA/NaI and DAMA/LIBRA Phase-I  have been released prior to publication of K18, they synthesized the data for the 4-5, 4-6, and 5-6 keV energy ranges for a fixed period and phase. However, in the latest data release for the DAMA/LIBRA experiment, the per-cycle modulation amplitudes for the 4-5 keV and 5-6 keV are provided in Figure 14 of \cite{Bernabei_2018}. We use this data (as given in Table ~\ref{tab:data_meancycle}) for our analysis.

While K18 analyzed the per-cycle data from DAMA/NaI and the first phase of the DAMA/LIBRA experiment,  the data analyzed here includes the  per-cycle modulation amplitude data from the first and second phase of the DAMA/LIBRA experiment that began in September 2003. We analyze data from the 1-2, 2-3, 3-4, 4-5 and 5-6 keV energy ranges. Values for the per-cycle amplitudes in the 2-6 and 4-6 keV ranges have not been provided by the DAMA collaboration in their 2018 data release, and hence have not been analyzed for the DAMA/LIBRA phase.
We also analyze the per-cycle modulation amplitudes for the 1-2 keV energy  range, which   have  been  released  only  for  DAMA/LIBRA  Phase-II and are not available for the DAMA/NaI phase  or DAMA/LIBRA Phase-I.

\begin{table*}[]
   \centering
   \resizebox{1.05\textwidth}{!}{
    \begin{tabular}{|c|c|c|c|c|c|c|}
        \cline{1-7}
        DAMA/LIBRA & $t$ & $S_m$ (1-2 keV) & $S_m$ (2-3 keV) & $S_m$ (3-4 keV) & $S_m$ (4-5 keV) & $S_m$ (5-6 keV)\\
         & ($year$) & ($/kg/day/keV$) & ($/kg/day/keV$) & ($/kg/day/keV$) & ($/kg/day/keV$) & ($/kg/day/keV$)\\
        \cline{1-7}
         Phase-I & & & & & & \\
        Cycle 1 & 1 & - & 0.0184 $\pm$ 0.0088 & 0.0361 $\pm$ 0.0082 &  0.0053 $\pm$ 0.0067 & -0.0049 $\pm$ 0.0067 \\
        Cycle 2 & 2 & - & 0.0137 $\pm$ 0.0101 & 0.0180 $\pm$ 0.0096 &  0.0093 $\pm$ 0.0071 & 0.0004 $\pm$ 0.0071 \\
        Cycle 3 & 3 & - & 0.0203 $\pm$ 0.0086 & 0.0073 $\pm$ 0.0089 &  0.0191 $\pm$ 0.0069 & -0.0003 $\pm$ 0.0069 \\
        Cycle 4 & 4 & - & 0.0284 $\pm$ 0.0079 & 0.0052 $\pm$ 0.0080 &  0.0091 $\pm$ 0.0063 & 0.0019 $\pm$ 0.0063 \\
        Cycle 5 & 5 & - & 0.0179 $\pm$ 0.0071 & 0.0033 $\pm$ 0.0071 &  0.0051 $\pm$ 0.0058 & 0.0076 $\pm$ 0.0058 \\
        Cycle 6 & 6 & - & 0.0195 $\pm$ 0.0073 & 0.0099 $\pm$ 0.0069 &  -0.0009 $\pm$ 0.0055 & 0.0139 $\pm$ 0.0055 \\
        Cycle 7 & 7 & - & 0.0224 $\pm$ 0.0067 & 0.0118 $\pm$ 0.0068 &  0.0022 $\pm$ 0.0050 & 0.0035 $\pm$ 0.0050 \\
        \cline{1-7}
        Phase-II & & & & & & \\
        Cycle 1 & 9 & 0.0323 $\pm$ 0.0074 & 0.0202 $\pm$ 0.0064 & 0.0212 $\pm$ 0.0057 &  0.0012 $\pm$ 0.0046 & 0.0076 $\pm$ 0.0046 \\
        Cycle 2 & 10 & 0.0147 $\pm$ 0.0081 & 0.0120 $\pm$ 0.0075 & 0.0162 $\pm$ 0.0064 & 0.0075 $\pm$ 0.0050 & 0.0079 $\pm$ 0.0050 \\
        Cycle 3 & 11 & 0.0232 $\pm$ 0.0080 & 0.0161 $\pm$ 0.0074 & 0.0076 $\pm$ 0.0062 & 0.0107 $\pm$ 0.0048 & 0.0034 $\pm$ 0.0048 \\
        Cycle 4 & 12 & 0.0356 $\pm$ 0.0081 & 0.0199 $\pm$ 0.0073 & 0.0108 $\pm$ 0.0061 & 0.0093 $\pm$ 0.0049 & 0.0026 $\pm$ 0.0049 \\
        Cycle 5 & 13 & 0.0095 $\pm$ 0.0084 & 0.0139 $\pm$ 0.0072 & 0.0121 $\pm$ 0.0063 & 0.0033 $\pm$ 0.0047 & 0.0069 $\pm$ 0.0047 \\
        Cycle 6 & 14 & 0.0049 $\pm$ 0.0077 & 0.0108 $\pm$ 0.0073 & 0.0079 $\pm$ 0.0061 & 0.0109 $\pm$ 0.0045 & 0.0054 $\pm$ 0.0045 \\

        \cline{1-7}    
    \end{tabular}}

    \caption{The per-cycle modulation amplitudes as determined by the DAMA collaboration, taken from Figure 36 of Ref \cite{BERNABEI2020103810}. The time axis corresponds to each annual cycle spanning over 14 years of the DAMA/LIBRA experiment, since the beginning of the DAMA/LIBRA Phase-I in September, 2003.} 
    \label{tab:data_meancycle}
\end{table*}

\begin{figure*}
        \centering
        \includegraphics[width=\textwidth]{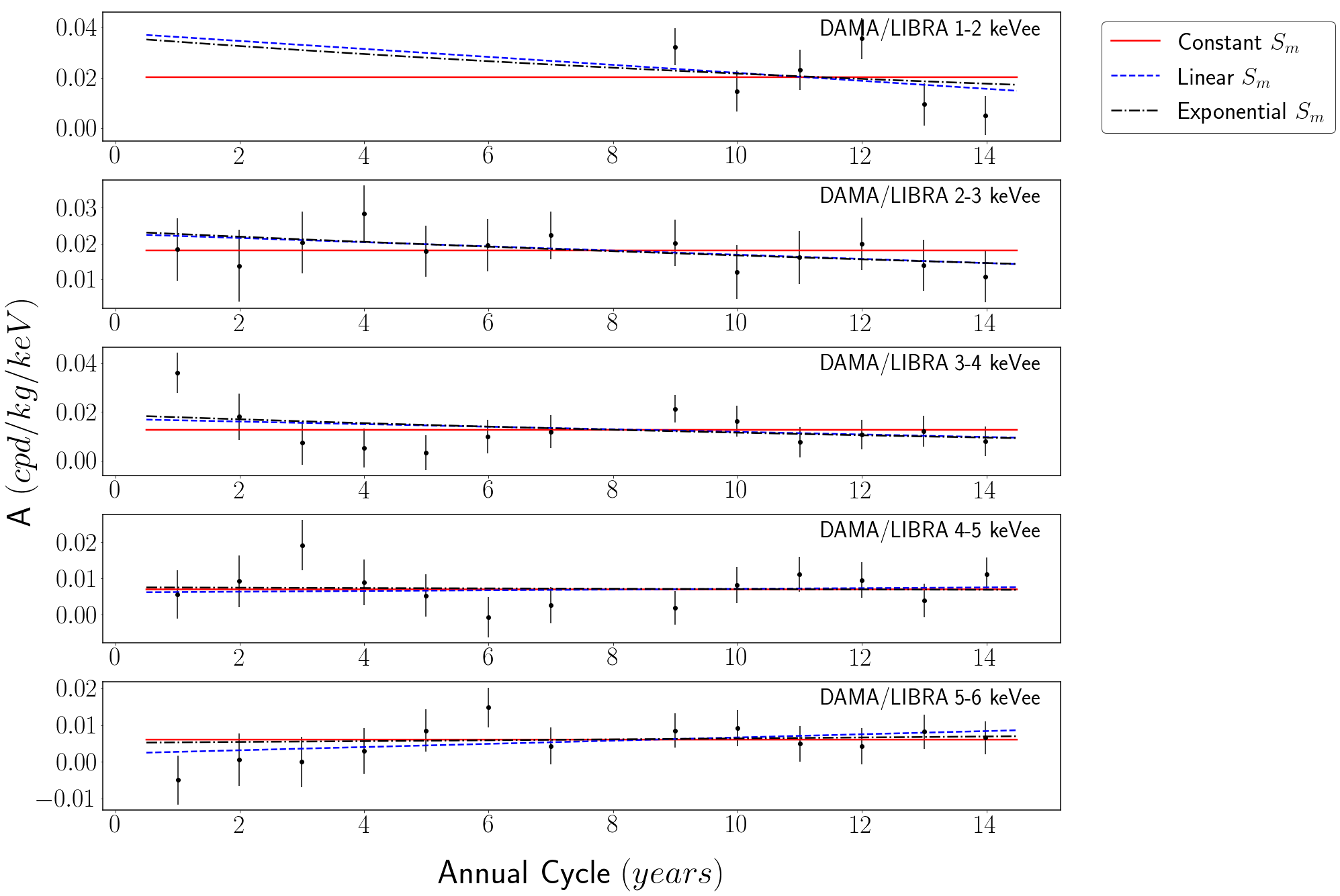}
        \caption{The best-fit curves for the constant, linear and exponential time-dependence of the modulation amplitude are plotted over the per-cycle mean amplitude values. The data for the 1-2 keV energy range has been released only for DAMA/LIBRA Phase-II. The x-axis represents each annual cycle spanning over the 14 years of the DAMA/LIBRA experiment, since the beginning of the DAMA/LIBRA Phase-I in September, 2003 \cite{BERNABEI2020103810}.}
        \label{fig:mean_value_plots}
\end{figure*}

 The  constant time-independent, linearly time-dependent and exponentially time-independent models for $S_m$ used in this analysis are given in Equations \ref{eq:mean_const}, \ref{eq:mean_linear} and \ref{eq:mean_exp}. The Bayesian priors used for the constant ($A_0$), linear ($m$, $c$), and exponential ($A$, $B$) model  parameters  remain the same as before and shown in Table \ref{table:bayes-priors1}.
  The best-fit plots of this analysis are given in Figure \ref{fig:mean_value_plots}. The best-ft values obtained from the posterior samples provided by {\tt Nestle} are shown in Table~\ref{tab:best-fit_mean}. 
  The $\chi^2$/DOF values and the Bayes factors have been listed in Columns III, and IV of Table \ref{table:results-2} respectively for the three models of $S_m$.   Unlike K18, we find that  $\chi^2$/dof estimates is close to 1 for all the time-varying models.  Furthermore, similar to our results for the residual count rates, we find that the  Bayes Factors  are close to 1. Therefore, we do not find any evidence  for a linear or exponential dependence of the (DAMA provided) modulation amplitudes  starting from the 2003 phase of DAMA/LIBRA.

\begin{table}[]
    \centering
    \resizebox{\textwidth}{!}{
    \begin{tabular}{|l l|c|c|c|c|c|}
    \cline{1-7}
        & & 1-2 keV & 2-3 keV & 3-4 keV & 4-5 keV & 5-6 keV \\
    \cline{1-7}
        Constant & $A_0$ $[10^{-2} dru]$ & $2.04 \pm 0.33$ & $1.80 \pm 0.21$ & $1.26 \pm 0.19$ & $0.70 \pm 0.15$ & $0.61 \pm 0.15$\\
    \cline{1-7}
        Linear & $m$ $[10^{-4} dru/year]$ & $15.92 \pm 9.08$ & $-5.78 \pm 5.38$ & $-5.35 \pm 4.87$ & $0.93 \pm 3.72$ & $4.38 \pm 3.74$\\
         & $c$ $[10^{-2} dru]$ & $3.79 \pm 1.01$& $2.27 \pm 0.47$ & $1.71 \pm 0.45$ & $0.63 \pm 0.35$ & $0.23 \pm 0.35$\\
    \cline{1-7}
        Exponential & $A$ $[10^{-2} dru]$ & $3.62 \pm 0.98$ & $2.36 \pm 0.52$ & $1.88 \pm 0.49$ & $0.76 \pm 0.32$ & $0.52 \pm 0.23$ \\
         & $B$ $[10^{-2} /year]$ & $5.11 \pm 3.01$ & $3.49 \pm 2.79$ & $4.92 \pm 3.23$ & $0.62 \pm 4.79$ & $-1.89 \pm 4.47$\\
    \cline{1-7}
    \end{tabular}}
    \caption{The best-fit values for the per-cycle modulation amplitudes as given by Equations \ref{eq:mean_const}, \ref{eq:mean_linear}, and \ref{eq:mean_exp}. The analysis and comparison of these models is described in Section \ref{sec:mean-value}. $dru$ corresponds to Differential Rate Unit, which is equal to 1 $[/kg/day/keV]$.} 
    \label{tab:best-fit_mean}
\end{table}

\begin{table*}[]
\centering

\begin{tabular}{|c|c|c|c|}
\cline{1-4}
Data & Models & $\chi^2/DOF$ & Bayes Factors \\ 
\cline{1-4}
DAMA/LIBRA Phase-II: & Constant & $12.5/5$ & - \\
1-2 keV & Linear & $7.6/4$ & $0.11$\\
 & Exponential & $8.2/4$ & $1.83$\\
\cline{1-4}

DAMA/LIBRA Phase-I and Phase-II: & Constant  & $4.6/12$ & - \\
2-3 keV & Linear  & $3.4/11$ & $0.02$\\
 & Exponential  & $3.5/11$ & $0.78$\\ 
 \cline{1-4}
 
DAMA/LIBRA Phase-I and Phase-II: & Constant  & $15.6/12$ & -\\
3-4 keV & Linear  & $14.4/11$ & $0.02$\\
 & Exponential  & $14.2/11$ & $1.34$\\ 
 \cline{1-4}

DAMA/LIBRA Phase-I and Phase-II: & Constant  & $9.7/12$ & - \\
4-5 keV & Linear  & $9.7/11$ & $0.01$\\
 & Exponential  & $9.7/11$ & $0.69$ \\ 
 \cline{1-4}

DAMA/LIBRA Phase-I and Phase-II: & Constant  & $8.2/12$ & - \\
5-6 keV & Linear  & $6.8/11$ & $0.02$\\
 & Exponential  & $7.2/11$ & $0.66$\\ 
 \cline{1-4}

\end{tabular}
\caption{Summary of model comparison results for different models of the per-cycle modulation amplitudes $S_m$ discussed in Section \ref{sec:mean-value}. \rthis{The $\chi^2$ values are calculated from the $\chi^2$ statistic by comparing the best-fit model to the data, and $DOF$ is taken as the difference in number of data points and the number of parameters.} The Bayes Factors for the linearly and exponentially varying modulation amplitudes have been calculated using data released by the DAMA collaboration for the first and second phase of the DAMA/LIBRA experiment \cite{Bernabei_2018}. The constant amplitude model is taken as the null hypothesis. We do not find any evidence for time modulation of the amplitudes from this analysis.}
\label{table:results-2}

\end{table*}

\section{Conclusions}
\label{sec:conclusions}
Around three years ago, K18 analyzed the DAMA annual modulation data collected over 14 cycles (starting from the DAMA/NaI phase in 1995) in order to discern a possible time dependence in the modulation amplitude. Their results indicate that all time varying amplitudes are preferred over a constant amplitude by upto 2-3$\sigma$
in almost all the energy ranges they analyzed. If the DAMA annual modulation is caused by astrophysical dark matter, one would not expect the modulation amplitude to change.

Motivated by these considerations, we revisit  this issue using the latest DAMA data.  We first re-analyze the modulation amplitudes used in K18 by carrying out Bayesian model comparison of the linear and exponential time-varying amplitudes versus the constant amplitude. Our results for this re-analysis can be found in Table~\ref{tab:kelso_bayes}.
We find that  although none of the models show a decisive evidence for any linear or exponential variation, the data in 2-5 keV and 2-6 keV energy intervals show Bayes factors of around 11 and 20, respectively, pointing to moderate evidence for a time-variation.

We then update this analysis using both the cumulative residual event rates as well as 
 the updated per cycle modulation amplitude provided by the DAMA collaboration in their latest paper~\cite{Bernabei_2018} starting from the DAMA/LIBRA phase  in 2003.  The analysis of   residual event rate was done by an augmented fit, consisting of a varying  modulation amplitude in addition to the cosine term.

Our results for the Bayesian model comparison using the single-hit residual rates and the per-cycle modulation amplitudes  are summarized in Table~\ref{table:results-1} and Table~\ref{table:results-2}, respectively. Using both the datasets, we find that the  Bayes factors for all the time-varying amplitudes with respect to the constant amplitude are less than or close to one.  Therefore, we do not find any evidence for a decreasing modulation amplitude as hinted from the K18 results for the DAMA/LIBRA phase of the DAMA experiment.

To promote transparency in data analysis and enable the reader to reproduce these results, we have uploaded all our codes datasets used in this analysis at \url{https://github.com/srinixbhagvati/Time-dependence-of-DAMA-LIBRA-modulation-amplitude}

\bibliography{main}
\end{document}